\documentclass[reprint,superscriptaddress,amsmath,amsfonts,amssymb,aps,prl,showkeys]{revtex4-1}

\newcommand{\blue}[1]{{\color{black}{#1}}}

\usepackage{bm}% bold math
\usepackage{xcolor}
\usepackage{graphicx}
\usepackage{amsmath}
\usepackage{amssymb}

\usepackage{hyperref}
\hypersetup{
  breaklinks = {true},
  citecolor = {blue},
  colorlinks = {true},
  linkcolor = {red},
  urlcolor={blue},
}

\graphicspath{{../Figs/}}

\begin{document}

\title{\blue{Low-symmetry topological materials for large charge-to-spin interconversion: the case of transition metal dichalcogenide monolayers}}

\author{Marc Vila}
\affiliation{Catalan Institute of Nanoscience and Nanotechnology (ICN2), CSIC and BIST, Campus UAB, Bellaterra, 08193 Barcelona, Spain}
\affiliation{Department of Physics, Universitat Aut\`onoma de Barcelona, Campus UAB, Bellaterra, 08193 Barcelona, Spain}

\author{Chuang-Han Hsu}
\affiliation{Department of Electrical and Computer Engineering, National University of Singapore, Singapore 117576, Singapore}
\affiliation{Centre for Advanced 2D Materials and Graphene Research Centre, National University of Singapore, Singapore 117546, Singapore}

\author{Jose H. Garcia}
\affiliation{Catalan Institute of Nanoscience and Nanotechnology (ICN2), CSIC and BIST, Campus UAB, Bellaterra, 08193 Barcelona, Spain}

\author{L. Antonio Ben\'itez}
\affiliation{Catalan Institute of Nanoscience and Nanotechnology (ICN2), CSIC and BIST, Campus UAB, Bellaterra, 08193 Barcelona, Spain}
\affiliation{Department of Physics, Universitat Aut\`onoma de Barcelona, Campus UAB, Bellaterra, 08193 Barcelona, Spain}
\affiliation{Department of Physics, Massachusetts Institute of Technology, Cambridge, MA, USA}

\author{Xavier Waintal}
\affiliation{Univ. Grenoble Alpes, CEA, IRIG-PHELIQS, 38000 Grenoble, France}

\author{Sergio O. Valenzuela}
\affiliation{Catalan Institute of Nanoscience and Nanotechnology (ICN2), CSIC and BIST, Campus UAB, Bellaterra, 08193 Barcelona, Spain}
\affiliation{ICREA--Instituci\'o Catalana de Recerca i Estudis Avan\c{c}ats, 08010 Barcelona, Spain}

\author{Vitor M. Pereira}
\affiliation{Centre for Advanced 2D Materials and Graphene Research Centre, National University of Singapore, Singapore 117546, Singapore}

\author{Stephan Roche}
\affiliation{Catalan Institute of Nanoscience and Nanotechnology (ICN2), CSIC and BIST, Campus UAB, Bellaterra, 08193 Barcelona, Spain}
\affiliation{ICREA--Instituci\'o Catalana de Recerca i Estudis Avan\c{c}ats, 08010 Barcelona, Spain}

\begin{abstract}
The spin polarization induced by the spin Hall effect (SHE) in thin films 
typically points out of the plane. This is rooted on the specific symmetries of traditionally studied systems, not in a fundamental constraint. Recently, experiments on few-layer ${\rm MoTe}_2$ and ${\rm WTe}_2$ showed that the reduced symmetry of these strong spin-orbit coupling materials enables a new form of {\it canted} spin Hall effect, characterized by concurrent in-plane and out-of-plane spin polarizations. Here, through quantum transport calculations on realistic device geometries, including disorder, we predict a very large gate-tunable SHE figure of merit $\lambda_s\theta_{xy}\sim 1\text{--}50$ nm in ${\rm MoTe}_2$ and ${\rm WTe}_2$ monolayers that significantly exceeds values of conventional SHE materials. This stems from a concurrent long spin diffusion length ($\lambda_s$) and charge-to-spin interconversion efficiency as large as $\theta_{xy} \approx 80$\%, originating from momentum-invariant (persistent) spin textures together with large spin Berry curvature along the Fermi contour, respectively. \blue{Generalization to other materials and specific guidelines for unambiguous experimental confirmation are proposed, paving the way towards exploiting such phenomena in spintronic devices. These findings vividly emphasize how crystal symmetry and electronic topology can govern the intrinsic SHE and spin relaxation, and how they may be exploited to broaden the range and efficiency of spintronic materials and functionalities}.
\end{abstract}

\maketitle

\section{Introduction}
Unconventional manifestations of spin-orbit coupling (SOC) are rapidly extending the ability to generate, control and carry spin polarization for applications of spin transport or spin-driven magnetic torques beyond conventional spintronic materials \cite{Han2018,Manchon2019,Dolui2020,Bonell2020}.
Topological materials form a natural family to scrutinize in this regard: their key features often derive from a large SOC combined with band inversions, and their topologically protected surface states may prove instrumental to enable coherent, dissipationless spin currents over long-distances \cite{Fan2016,Kononov2020}. 
3D Weyl semimetals (WSM) are defined by the presence of band degeneracy points near the Fermi energy ($E_F$) with local linear dispersion in all directions \cite{Burkov2011,Wan2011}. Layered transition metal dichalcogenides (TMDs) in the 1T$^\prime$ ($P2_{1}/m$) or 1T$_\text{d}$ ($Pmn2_{1}$) phases accommodate the interesting class of WSM candidates MX$_2$ (M\,=\,Mo,\,W; X\,=\,S,\,Se,\,Te), which have been advanced as platforms for realizing exotic phenomena such as topological superconductivity \cite{Qi2016,Fatemi2018,Sajadi922,Chiu2020}, non-linear Hall effect \cite{Ma2019, Kang2019, Xu2018, Zhang2018,Singh2020}, anisotropic spin Hall transport \cite{Zhou2019} or out-of-plane spin-orbit torque \cite{MacNeill2017, Li2018}. When thinned towards the monolayer limit, they transition from the type-II WSM bulk phase to the quantum spin Hall regime \cite{Kane2005, Kane2005QSH, Qian2014, Soluyanov2015,Sun2015, Tang2017, Fei2017, Jia2017, Chen2018, Peng2017, Wu2018, Shi2019,Xu613,Jiang2017, Li2017} with strain-tunable topological gap \cite{Zhao2020strain}. 

Recently, large charge-to-spin interconversion (CSI) generated by the spin Hall effect (SHE) has been reported in multilayers of MoTe$_2$ and WTe$_2$ \cite{Safeer2019, Song2020, Zhao2020, Zhao2020AM}. The CSI efficiency is quantified in terms of the spin Hall angle (SHA, $\theta_{xy}$), which indicates what fraction of a driving charge current ($\bm{J}_c$) can be converted into spin current ($\bm{J}_s^\alpha$); $\theta_{xy}$ depends on the magnitude of SOC and is typically no more than a few percent at room temperature in heavy metals \cite{Sinova2015}. In traditional SHE materials, the stronger the SOC, the shorter the spin diffusion length ($\lambda_s$) is; consequently, achieving long $\lambda_s$ concurrently with large SHA is a long-standing challenge for spintronics. To date, the best tradeoff obtained with heavy metals amounts to $\lambda_s\theta_{xy}\sim 0.1\text{--}0.2$ nm \cite{Sinova2015,Hoffmann2013, Isasa2015,Laczkowski2017,Sagasta2018,Sayed2021}.

Interestingly, hints of unconventional SHE have been detected in 1T$^\prime$ \cite{Safeer2019} and 1T$_\text{d}$ phases of MoTe$_2$ multilayers \cite{Song2020}, 
characterized by spin currents carrying spins ($\bm{S}$) collinear with the charge current, which is unique so far. This is possible because, in contrast to bulk crystals, the absence of the glide mirror symmetry in few-layer slabs allows for additional non-zero components of the spin Hall conductivity (SHC) tensor \cite{Seemann2015}, thus breaking away from the traditional constraint imposing a stringent right-hand rule $\bm{J}_c \perp \bm{J}_s \perp \bm{S}$). Furthermore, a remarkably long spin diffusion length ($\lambda_s$) has been reported in MoTe$_2$ \cite{Song2020}, although its actual magnitude is under debate \cite{Safeer2019, Zhao2020, Zhao2020AM}. These experimental developments call for an understanding of the mechanisms governing spin dynamics in such systems, especially the large CSI efficiency $\theta_{xy}$ and what enables large $\lambda_s$, given that these are traditionally \emph{anti}-correlated quantities.

\blue{While topological materials could be expected to display large SHE efficiency, primarily driven by a strong SOC and a large intrinsic SHC, here we show that, in addition, the reduced symmetry plays a central role by enabling otherwise-forbidden persistent (i.e., $\bm{k}$-invariant) spin textures (PST) that sustain unusually large spin diffusion lengths.}
Specifically, we show that unique symmetry-induced spin textures of electronic states in MoTe$_2$ and WTe$_2$ monolayers yield a {\it giant} canted SHE where the spin current polarization lies in the $yz$ plane. CSI efficiencies can be as high as $80\%$ and values of $\lambda_s$ in the range 10--100 nm, up to one order of magnitude larger than in heavy metals with similar spin Hall efficiency \cite{Sinova2015}. We unveil that these large values arise from the interplay of a PST, confirmed by density functional theory (DFT), and a large spin Berry curvature (SBC), stemming from the band inversion and hybridization near $E_F$ that underlies the nontrivial topology of these monolayers. Moreover, all spin transport characteristics are gate-tunable, being maximal near the band edge and allowing a CSI figure of merit of up to $\lambda_s\theta_{xy} \sim 50$ nm. 
Our findings hinge on DFT, symmetry considerations and a purpose-built effective tight-binding model deployed in spintronic simulations using state-of-the-art quantum transport methodologies \cite{Fan2021}.
\blue{Importantly, we discuss how similarly large spintronic figures of merit can be anticipated, based on the same key physical ingredients, among several other classes of materials that would be interesting to explore. This paves the way to uncovering systems endowed with both large $\theta_{xy}$ and large $\lambda_s$, breaking free from a challenging constraint that has hindered spintronic applications.}

\section{Theoretical model}

We computed the bandstructures of 1T$_\text{d}$-derived monolayers of MoTe$_2$ and WTe$_2$ within DFT (see supplemental material \cite{Suppmat}). Effective Hamiltonians based on maximally localized Wannier functions were subsequently extracted, allowing straightforward computation of the SHC and spin textures with no intervening approximations. Yet, such Hamiltonian is still too complex to be efficiently deployed in large-scale transport calculations on system sizes involving millions of unit cells. 
We therefore built a $\bm{k}\cdot\bm{p}$ Hamiltonian to describe the two conduction and two valence bands nearest $E_F$ which, at the $\Gamma$ point, transform according to the representations $B_u$ (valence) and $A_g$ (conduction) of the $C_{2\text{h}}$ point group \cite{Qian2014} (see our supplementary\cite{Suppmat} ``Remarks on the 4-band model" for further discussion). Extension of the symmetry-allowed $\bm{k}\cdot\bm{p}$ terms to the full Brillouin zone yields the following nearest-neighbor tight-binding representation \cite{Suppmat}:
\begin{align} \label{eq_H}
H &=\sum_{i,s}(\Delta + 4m_p+ \delta) c_{i,s}^\dagger c_{i,s} \nonumber  \\
&-\sum_{\langle 
ij\rangle,s} (m_p  + m_d)c_{i,s}^\dagger c_{j,s} \nonumber 
\\
&+\sum_{i,s}(\Delta  - 4m_d - \delta) d_{i,s}^\dagger d_{i,s} \nonumber \\
&-\sum_{\langle ij \rangle,s}  (m_p  - m_d)d_{i,s}^\dagger d_{j,s} \nonumber 
\\
&- \sum_{\langle ij\rangle,s} \frac{\beta}{2} (\hat{\bm{l}}_{ij}\cdot 
\hat{\bm{y}} ) \, c_{i,s}^\dagger d_{j,s} + \sum_{i,s} \eta  c_{i,s}^\dagger 
d_{i,s}  \nonumber 
\\
&-\sum_{\langle ij\rangle}\sum_{ss'} \frac{i}{2} (\bm{\Lambda}_{ss'}\times 
\hat{\bm{l}}_{ij})\cdot(\hat{\bm{y}}+\hat{\bm{z}} ) c_{i,s}^\dagger d_{j,s'} \nonumber
\\
&+\text{H.c}.
\end{align}
This is an effective 4-band Hamiltonian generated by two orbitals (plus spin) per unit cell on a rectangular lattice, one arising from the chalcogen $p_y$ states and the other from metal $d_{yz}$ orbitals, respectively associated with the $c_{i,s}$ and $d_{i,s}$ operators at each unit cell $i$ ($s$ labels the spin projection). 
The first four terms in Eq.~\eqref{eq_H} describe spin-degenerate valence and conduction bands with hopping amplitudes set by $m_p \pm m_d$, $\delta$ parameterizes the degree of band inversion at $\Gamma$, and a constant $\Delta$ is used to match the position of the conduction band and $E_F$ with those obtained by DFT. 
In the fifth term, $\beta$ accounts for the $x$--$y$ crystalline anisotropy, with $\hat{\bm{l}}_{ij}$ a unit vector pointing from site $i$ to $j$; the term $\propto \eta$ breaks inversion symmetry and determines, for example, whether we are describing a monolayer descended from a 1T$^\prime$ ($\eta = 0$) or 1T$_\text{d}$ bulk crystal.
The last term embodies the SOC, where $\bm{\Lambda} \equiv (\Lambda_x \sigma_x,-\Lambda_y \sigma_y, \Lambda_z \sigma_z)$, $\sigma_{x,y,z}$ are the spin Pauli matrices and $\hat{\bm{y}}$, $\hat{\bm{z}}$ are unit Cartesian vectors. The parameters are set by fitting the energy dispersion and spin texture to the ones obtained by DFT \cite{Suppmat}. 
To be specific, we henceforth concentrate on the case of MoTe$_2$, as it is the one where experiments have recently reported in-plane SHE \cite{Safeer2019, Song2020}. Nonetheless, because the Hamiltonian model \eqref{eq_H} captures equally well the case of WTe$_2$ and similar low-symmetry TMDs \cite{Suppmat}, qualitatively comparable results can be expected in those monolayers as well. Additionally, we found very similar results between the 1T$_\text{d}$ and 1T$^\prime$ phases and therefore we focus here on the former while the latter is reported in the supplemental material \cite{Suppmat}. Since the DFT calculations show that MoTe$_2$ is slightly $n$-doped, we favored the conduction band in the fits and will focus exclusively on cases where $E_F$ lies in the conduction band.

\section{Numerical Spin Dynamics Calculations}
\begin{figure}
\includegraphics[width=0.45 \textwidth]{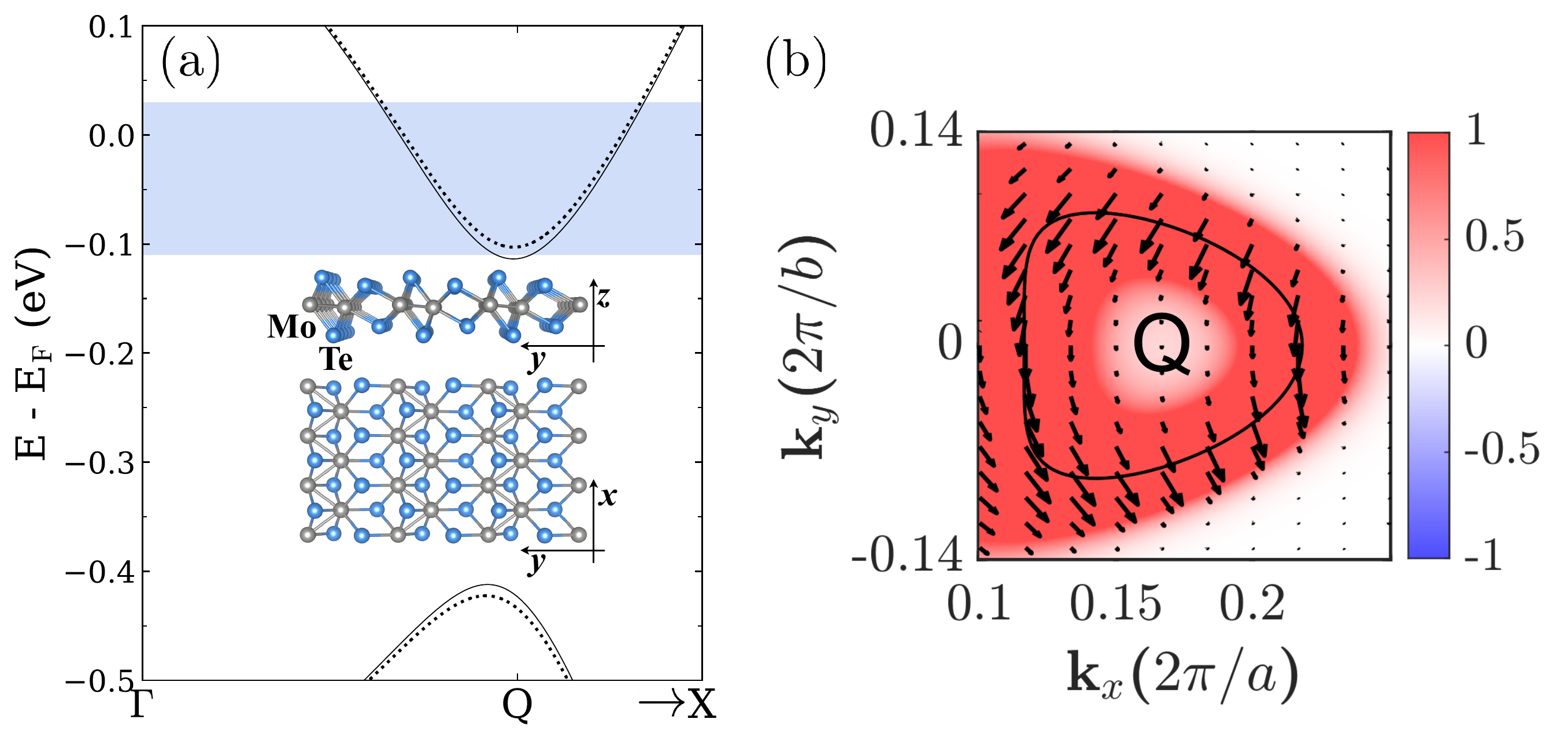}
\caption{(a) \blue{Close-up of the bandstructure near $E_F$ for the 4-band model (Eq. (\ref{eq_H})) of T$_\text{d}$-MoTe$_2$ monolayer near one of the time-reversal-symmetric $Q$ points}. The blue-shaded region indicates the energy range covered in the spin transport calculations. Inset: monolayer crystal structure. (b) Spin texture of one of the bands of the electron pocket near $\text{Q}$ at $E_F$ (Fermi-broadened with $T = 300$ K); the solid line marks the Fermi contour, arrows depict the in-plane spin projection and the color indicates the spin projection along $z$.} 
\label{fig_F1}
\end{figure}

In Fig.~\ref{fig_F1}a, \blue{we plot the band structure of the aforementioned 4-band model} near $E_F$ and one of the time-reversal-symmetric Q points, where the valence and conduction extrema occur. The small splitting arises from the small inversion-symmetry-breaking ($\eta\ne0$) that occurs in monolayers derived from the T$_\text{d}$ bulk structure. Since this work focuses on the scenario where $E_F$ lies in the conduction band, the figure shows a closeup of the conduction electron pockets; the full DFT bandstructure and the $\bm{k}{\,\cdot\,}\bm{p}$ fit are discussed in the supplemental material \cite{Suppmat}. Fig.~\ref{fig_F1}b shows the spin texture at the Fermi energy, $\langle s^\alpha\rangle_{E_\text{F}}$, with two crucial observations: the existence of an approximate \emph{persistent spin texture} through the whole Fermi contour \cite{Schliemann2003, Bernevig2006, Schliemann2017} and canted spins with $\langle s^y\rangle_{E_\text{F}} \gtrsim \langle s^z\rangle_{E_\text{F}} \gg \langle s^x\rangle_{E_\text{F}}$, consistent with prior studies \cite{Xu2018, Shi2019prb}.

\begin{figure}%[tb]
\includegraphics[width =0.45 \textwidth]{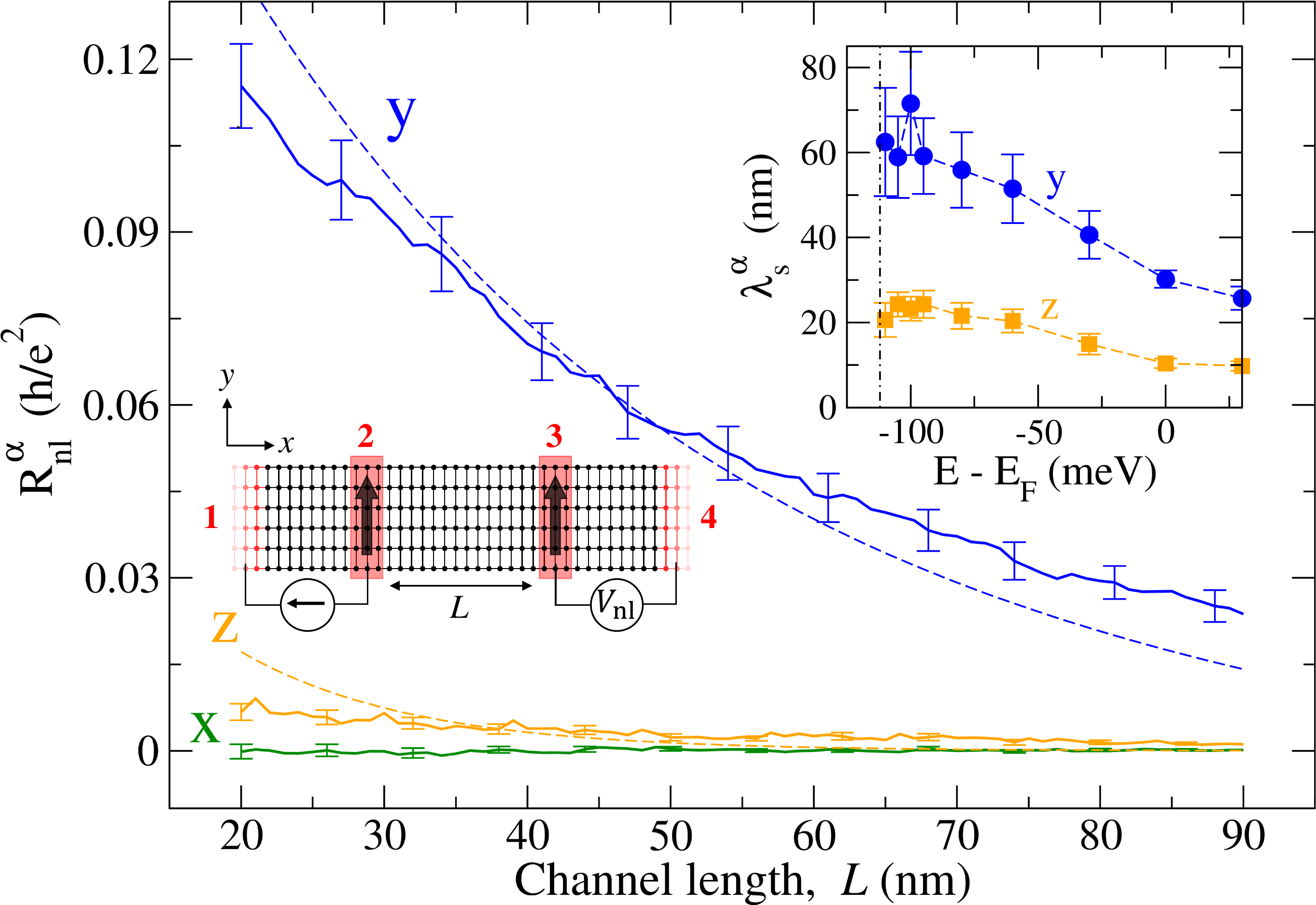}
\caption{$R_\text{nl}^\alpha$ (solid lines) against the channel length, 
$L$, for spins polarized along $x$, $y$ and $z$. Error bars result from 
averaging over 150 disorder configurations ($w$ = 50 nm). Dashed lines are fits to Eq. (10) in \cite{Suppmat}. Left inset: Scheme of 
the nonlocal spin valve. Black (red) regions denote the device (leads), with leads 2 and 3 being ferromagnetic. Current $I_0^\alpha$ flows from lead 2 to 1 and $V_\text{nl}$ is measured between leads 3 and 4. Right inset: Energy-dependence of $\lambda_s^{y,z}$. The dot-dashed line marks the conduction band minimum.} 
\label{fig_F2}
\end{figure}

We explored the spin transport properties using linear response theory and the Landauer-B\"{u}ttiker formalism as implemented in Kwant \cite{Groth2014}. We simulated the nonlocal spin valve illustrated in the inset of Fig.~\ref{fig_F2}, where contacts 2 and 3 are ferromagnetic (FM) to allow injection and detection of spin-polarized currents \cite{Johnson1985, Jedema2002, Fabian2007}: FM electrode 2 injects a spin-polarized current $I_0^\alpha$ with spins polarized along $\alpha \in \{x,y,z\}$; this creates a spin accumulation that diffuses along the channel and is detected as a nonlocal voltage $V_\textrm{nl}$ at electrodes 3-4, located a distance $L$ from the source and far from the path of charge current between electrodes 1-2. This effect is quantified by the  nonlocal resistance $R_\text{nl}^\alpha \equiv V_\text{nl}/I_0^\alpha$. 
The spin diffusion length for $\alpha$-pointing spins, $\lambda^{\alpha}_s$, is obtained from the decay of $R^{\alpha}_\text{nl}$ with $L$ in the diffusive regime (mean free path shorter than $L$). To ensure our results reflect the diffusive regime, we add Anderson disorder to the Hamiltonian and extract statistics only within the appropriate scaling region of the device conductance \cite{Vila2020, Suppmat}. 

Fig.~\ref{fig_F2} shows $R_\text{nl}^\alpha (L)$ for the three spin orientations at $E_F$. We see clear differences in the magnitude of the nonlocal signals and their relaxation distances for different orientations of the injected spins, ranging from tens of nanometers to the sub-nanometer scale. By fitting the length dependence of $R_\text{nl}^\alpha$ to the solution of the spin diffusion equations  (dashed lines in Fig.~\ref{fig_F2}) \cite{Vila2020}, we obtained $\lambda_s^{y} \approx 30$ nm and $\lambda_s^{z} \approx 10$ nm, while $\lambda_s^{x}$ has a negligible value. These values are comparable with strong-SOC metals such as Pt, $\beta$-W or $\beta$-Ta \cite{Sinova2015,Isasa2015}. 
It is significant that the spin diffusion lengths follow the trend $\lambda_s^{y} \gtrsim \lambda_s^{z} \gg \lambda_s^{x}$, in correspondence to that of the spin texture around the Fermi contour at equilibrium. The upper inset of Fig.~\ref{fig_F2} shows that this hierarchy holds over the entire range of energies analyzed, from $E = 30$ meV to the band edge at $\sim -110$ meV (we measure energies relative to $E_F$ of undoped MoTe$_2$). Details of how the (persistent) spin texture impacts $\lambda_s^{\alpha}$ and its scaling with energy are discussed in the supplemental material \cite{Suppmat}. Both $\lambda_s^{y}$ and $\lambda_s^{z}$ increase about threefold as $E_F$ moves towards the band edge (dot-dashed line).  Moreover, we found that $\lambda_s^{y}$ at $E_F = -140$ meV (in the band gap) increases up to $\simeq 156$ nm, while deeper into the gap ($E = -320$ meV) we see no decay in the spin signal, consistent with the onset of ballistic regime where spin is transported by topologically protected surface states \cite{Suppmat}. 

\begin{figure}%[tb]
\includegraphics[width =0.45\textwidth]{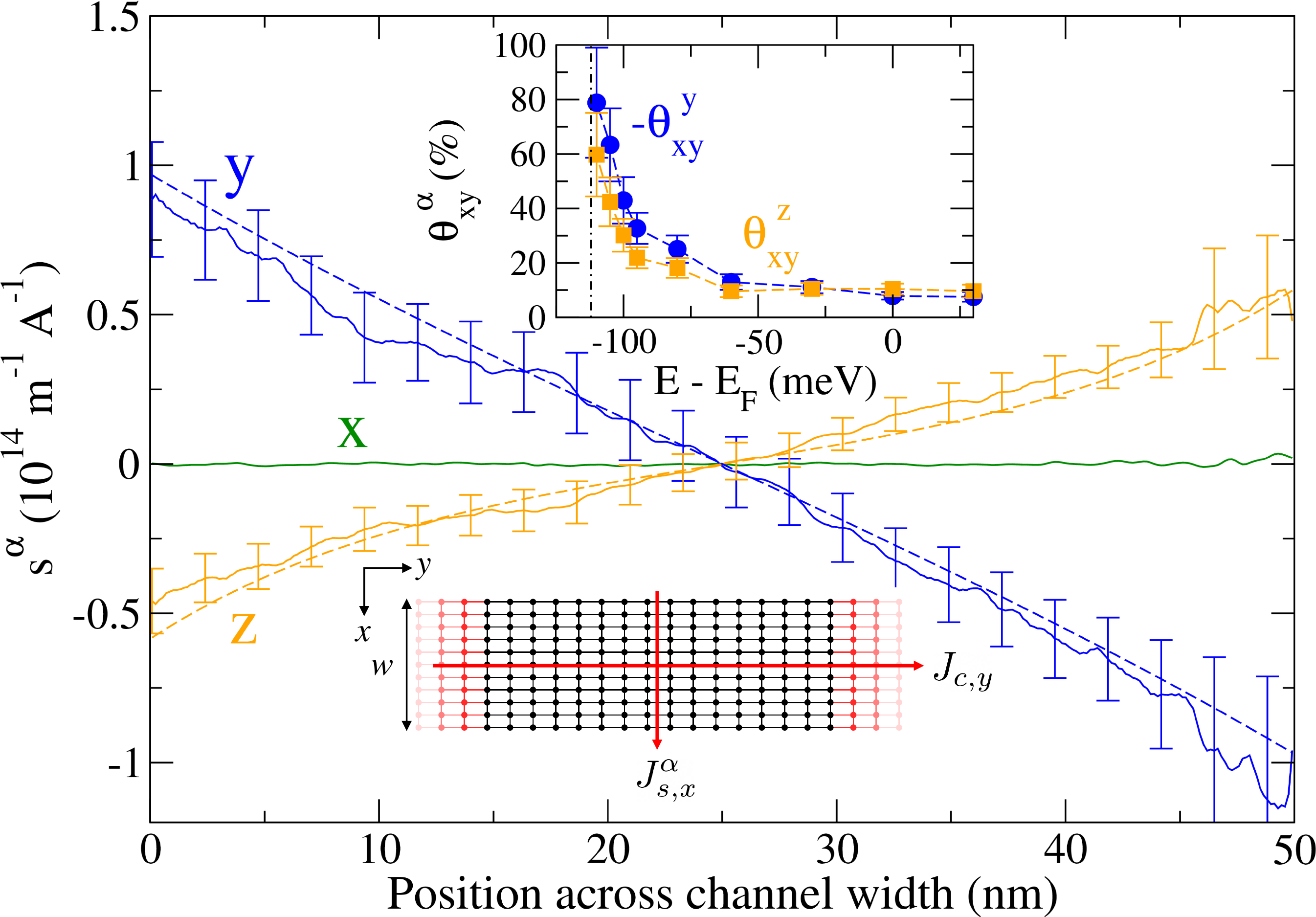}
\caption{Spin accumulation (solid lines) as a function of position across the channel width, of spins along $x$, $y$ and $z$. Error bars result from averaging over 200 disorder configurations ($w$ = 50 nm). Dashed lines are fits to Eq. (\ref{eq_Sdiff}). Bottom inset: Scheme of the two-terminal device, where a current flowing along $y$ creates a spin accumulation in the $x$ direction. Top inset: Energy dependence of the spin Hall angles, with the conduction band minimum marked by a dot-dashed line.}
\label{fig_F3}
\end{figure}

The SHE was investigated by computing the spin accumulation and analyzing its polarization, $s^\alpha$, along the three Cartesian directions $\alpha$. Charge current along $y$ generates a transverse spin current parallel to $x$ by the SHE, which results in spin accumulation at the open lateral boundaries. 
The efficiency of CSI is characterized by the SHA, defined as the ratio $\theta_{ij}^\alpha \equiv J_{s,i}^\alpha/J_{c,j}$, where $\hbar\bm{J}_s^\alpha/2e$ ($\bm{J}_c$) is the spin (charge) current density, $e$ the electron charge, and $i,j{\,\in\,}\{x,y\}$ denote the respective current directions. To numerically determine the SHA, we calculated the spin accumulation response function per unit of current applied to the lead, and fit it to the solution of the spin drift-diffusion equations: 
\begin{equation}\label{eq_Sdiff} 
\frac{s^\alpha (x)}{J_{c,y}} = - 
\frac{\theta_{xy}^{\alpha} \lambda_s^\alpha }{|e| 
D_s}\frac{\sinh(\frac{w-2x}{2 \lambda_s^{\alpha}}) 
}{\cosh(\frac{w}{2\lambda_s^{\alpha}})},
\end{equation} 
where $w$ is the device width and $D_s$ is the spin diffusion coefficient (see supplemental material \cite{Suppmat} for details).

Fig.~\ref{fig_F3} shows the averaged spin accumulation along the channel cross-section, $s^{\alpha}(x)$, for each spin orientation. In a typical SHE scenario, the electrical current, spin current, and the spin polarization are all mutually orthogonal (for this geometry, that would generate a finite $s^z$ only); however, we observe a nonzero $s^y$ as well due to the absence of a glide mirror symmetry in 1T$_\text{d}$-MoTe$_2$ monolayer. In fact, $|s^z|\sim|s^y|$, implying that the accumulated spins point \emph{obliquely} in the $yz$ plane, with significant projection parallel to the electrical current. Interestingly, note that the spin accumulation displays the hierarchy $s^y \gtrsim s^z \gg s^x$, echoing the trend seen above for the spin texture and spin diffusion lengths.

We determine $\theta_{xy}^\alpha$ by fitting the numerically calculated spin accumulation to Eq.~\eqref{eq_Sdiff}, using the values of $D_s$ extracted from the two-terminal conductance of this device and $\lambda_s^{\alpha}$ from Fig.~\ref{fig_F2} \cite{Suppmat}. The results are displayed in the inset of Fig.~\ref{fig_F3}. (We note that while the charge conductivity along $x$ and $y$ is slightly anisotropic, resulting in an equally anisotropic SHA,  $|\theta^\alpha_{xy}|$ and $|\theta^\alpha_{yx}|$ are still very similar \cite{Suppmat}.) At $E_F$, the SHA for spins pointing along $y$ and $z$ is $\approx 10\%$ (with opposite sign). Remarkably, both increase substantially when approaching the band edge, at which point $|\theta_{xy}^y|$ slightly overcomes $|\theta_{xy}^z|$ with values as large as $|\theta_{xy}^y| \approx 80\%$. We also computed the SHC and the SHA with the Kubo formula and obtain the same result both qualitatively and quantitatively (supplementary Fig.~11). The increase of $\theta_{xy}$ is attributed to hotspots of SBC near the bottom of the electron pockets \cite{Song2020, Xu2018} (supplementary Figs.~4--6), which directly determine the SHC/SHA magnitude \cite{Sinova2004, Tanaka2008, Zhou2019, Suppmat}.
Importantly, our combined results yield a CSI figure of merit $\lambda_s \theta_{xy}\sim 1\text{--}50$ nm, with the largest values attained at the band edge and for $y$-pointing spins. The upper limit exceeds that of traditional SOC materials (Pt, $\beta$-W, $\beta$-Ta or Au) for which $\lambda_s \theta_{xy}\sim 0.1-0.2$ nm \cite{Sinova2015,Isasa2015,Roy2017}, and is up to 2 to 3 times larger than that induced by proximity in graphene \cite{Safeer2019Gr, Benitez2020, Herling2020}. Such remarkable figure of merit stems from the combination of large SBC \emph{and} the persistent spin texture near the MoTe$_2$ band edges \cite{Suppmat}. These results represent the expected behavior in the monolayer limit of recent experiments performed on few-layer MoTe$_2$ \cite{Safeer2019, Song2020} and WTe$_2$ \cite{Zhao2020, Zhao2020AM}.

\section{Experimental detection}

\begin{figure*}%[tb]
\includegraphics[width=1.0\textwidth]{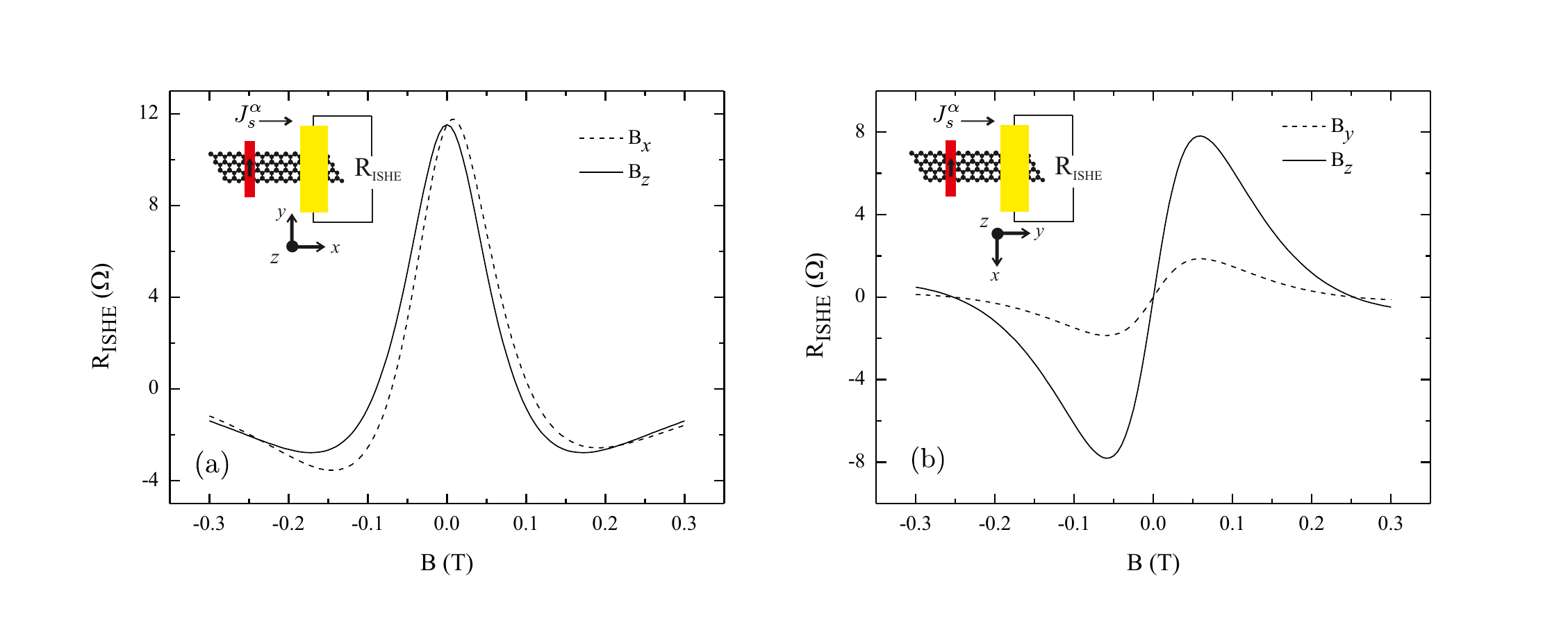}
\caption{Simulated response of the inverse SHE ($R_\mathrm{ISHE}$) to spin precession for two orientations of the TMD crystal (coordinate axes in the insets). The device geometry is shown in the insets, with the TMD depicted in yellow and the FM injector in red (magnetization indicated by an arrow). The polarization of the spin current reaching the TMD ($\bm{J}_s^\alpha$) is controlled externally with a magnetic field, $B$, oriented either along the graphene channel (dashed lines) or out-of-plane (solid lines). Typical experimental device dimensions were used in the simulation \cite{Suppmat}.} 
\label{fig_F4}
\end{figure*}

Such a peculiar spin response should become manifest in suitably designed nonlocal spin-precession experiments \cite{SaveroTorres2017, Safeer2019, Benitez2020,Cavill2020}. To probe this canted SHE, we propose the device concept pictured in the insets of Fig.~\ref{fig_F4}, which relies on the reciprocal/inverse SHE (ISHE) \cite{Sinova2015}. It consists of a Hall bar comprising a graphene channel and a transversely aligned monolayer TMD crystal. 
A non-equilibrium spin accumulation is induced in the graphene channel through a FM electrode whose magnetization direction determines that of the spin density injected into graphene underneath. This generates a pure-spin current that diffuses toward\,---\,and is absorbed by\,---\,the remote TMD. It is assumed that the spin current is absorbed by the TMD at its edge and continues to follow the diffusion direction, given that the spin resistance in the TMD is two orders of magnitude lower than in graphene for $\lambda_s^y = 30$ nm or $\lambda_s^z = 10$ nm \cite{Suppmat}. By ISHE, a transverse voltage $V_\mathrm{ISHE}$ develops on the TMD, which can be measured along its length as illustrated in Fig.~\ref{fig_F4}. In experiments, the diffusing spins can be controlled by spin precession in a non-collinear magnetic field {\it B}. To capture this situation, we generalized the Bloch diffusion equations to account for anisotropic spin diffusion and calculated $V_\mathrm{ISHE}(B)$ using the approach described in Ref. \cite{Benitez2020} (which accurately reproduces CSI in real devices).
Fig.~\ref{fig_F4} shows the precession response for two selected orientations of the TMD crystal in the limit of full absorption ($R_\mathrm{ISHE}\equiv V_\mathrm{ISHE}/I_0^y$) \cite{Suppmat}. We observe magnitudes of $R_\mathrm{ISHE}$ nearly three orders of magnitude larger than the values reported for graphene/TMDs \cite{Safeer2019Gr, Ghiasi2019, Benitez2020, Herling2020} and graphene/bulk-WSMs \cite{Safeer2019, Zhao2020}. This is a direct consequence of the extremely large SHA predicted here for MoTe$_2$ \cite{footnote}.

The essence of the experiment is that the precession response depends strongly on the crystal orientation. As evidenced in Fig.~\ref{fig_F2}, the spin relaxation in the TMD is anisotropic, and the CSI depends crucially on both the majority spin orientation and crystalline orientation.
In Fig.~\ref{fig_F4}(a), the TMD's crystallographic $y$-axis is transverse to the spin propagation. A magnetic field parallel to $z$ causes spins to precess in the graphene plane but, according to Fig.~\ref{fig_F3}, only the $y$ spin projection contributes to the ISHE signal with an efficiency of $|\theta_{xy}^y|$; $R_\text{ISHE}$ is symmetric with respect to the sign of $B$ because the magnetization at the FM injector is parallel to $y$, resulting in the maximum signal at $B=0$. When the field is parallel to $x$, the spins acquire a $z$ component (in addition to that in $y$), which is asymmetric with respect to $B$ and adds a contribution to the ISHE with an efficiency of $|\theta_{xy}^z|$; because $|\theta^y_{xy}| \gtrsim |\theta^z_{xy}|$, the signal remains roughly symmetric.
In Fig.~\ref{fig_F4}(b), the crystallographic $y$-axis is parallel to the spin propagation. As the $y$ and $z$ directions are now orthogonal to the FM magnetization, the lineshapes are antisymmetric. The signal is zero at $B=0$ and, by sweeping $B$ from negative to positive along $z$ ($x$), the spin component along $y$ ($z$) changes sign. Therefore, two combined observations in this proposed experiment represent a ``smoking gun'' demonstration of the intrinsic canted SHE predicted in this work: (i) $R_\mathrm{ISHE}(B)$ should display a different lineshape under different field orientations for a fixed TMD crystal; (ii) rotation of the crystal converts the lineshapes from predominantly symmetric to antisymmetric. 

\blue{
\section{Generalization to other systems}

Our numerical calculations show that the 1T$_\text{d}$ and 1T$^\prime$ phases of monolayer TMDs exhibit a canted spin Hall effect with large spin diffusion lengths and spin Hall angles. We precisely pinpoint this to the concurrence of a PST (which naturally enhances $\lambda_s$) and hotspots of SBC near the band edge. Additionally, the reduced symmetry has two fundamental consequences: it allows the extra non-zero components in the SHC tensor (hence a canted SHE) and it allows the PST (hence a large $\lambda_s$). 

It is important to note that, though we focused here on the MoTe$_2$ family in order to \emph{quantitatively} demonstrate all the above features, the concurrence of PST and large SBC may now be expected in precisely identifiable space groups (SG). 
On the one hand, Tao and Tsymbal \cite{Tao2018} have recently shown that robust PSTs (i.e., not relying on fine-tuned parameters \cite{Schliemann2017}) likely arise in crystals belonging to some nonsymmorphic space groups; this symmetry-based approach can be extended to enumerate all compatible crystal families and the features of their allowed PSTs. 
On the other hand, crystal families compatible with SHC components other than $\sigma^{z}_{xy}$ have been also enumerated \cite{Seemann2015, Zhang2021}, thereby identifying all the possible material platforms for canted, multi-component SHE. 
Finally, the magnitude of intrinsic SHE depends directly on the strength of the SOC and on the existence of non-zero SBC (see our supplementary ``Remarks on the spin Berry curvature''). Although both might be serendipitously large in some materials, topological materials are preferred for a targeted pursuit. This is because SOC combined with the underlying band inversions of, e.g. topological insulators or topological semimetals, invariably leads to SBC hotspots \cite{Kane2005QSH, Sun2016, Khang2018}. Since space groups have also been recently classified according to their compatibility with different topological phases \cite{Zhang2019, Vergniory2019}, one may systematically select those topological classes whose symmetries are simultaneously compatible with PST and canted SHC.

To be more specific, we provide materials that satisfy the qualities mentioned above, thus making them potential candidates for showing large CSI figures of merit. We focused on materials from the space groups reported in Ref. \cite{Tao2018} as they present PST. Although they only comprise orthorhombic crystals with no inversion symmetry, we note that extending the symmetry analysis of Ref. \cite{Tao2018} to other crystal systems may provide greater number of potential SHE materials with PST. We searched these space groups in the Topological Material Database \cite{Bradlyn2017, Vergniory2019, Vergniory2021, webs}, and among the several topological insulators and semimetals, we searched which of those had already been experimentally characterized. At least three topological materials were found: the type-II Weyl semimetal Ta$_3$S$_2$ (SG 39) \cite{Chang2016TAS} and AuSn$_4$ (SG 41), which is a topological nodal-line semimetal \cite{Shen2020}. According to the symmetry requirements for multi-component SHE \cite{Seemann2015}, the space groups from Ref. \cite{Tao2018} in their bulk form cannot host a canted SHE. However, these restrictions are lifted in systems with glide mirrors and screw axis with vertical translation when going to the monolayer limit, as in 1T$_\text{d}$-MoTe$_2$. This is indeed the case for TaIrTe$_4$ (SG 31), a van der Waals material being a Weyl semimetal in three-dimensions but a topological insulator in monolayer form \cite{Haubold2017, Belopolski2017, Liu2017}. Overall, we have identified Ta$_3$S$_2$, AuSn$_4$ and TaIrTe$_4$ as potential compounds for large CSI, with monolayer TaIrTe$_4$ also being compatible with multi-component SHE.
 
\section{Potential applications}

Discovering materials with largest $\lambda_s\theta_{xy}$ has been a long-standing challenge. This is partly because traditional understanding of SHE and spin diffusion posits that, while large SOC boosts the generation of spin current via SHE (quantified by $\theta_{xy}$), it detrimentally reduces the spin diffusion length \cite{Sinova2015}. Our extensive and realistic quantum simulations \emph{directly} demonstrate, for the first time, that materials hosting low-symmetry-enabled PST (even if only approximate) break free from that adverse compromise while, at the same time, displaying a new canted SHE, which greatly increases the geometrical flexibility of possible SHE-based devices. 
These results come at a time of impressive achievements in using 2D materials to carry spin currents over long distances and controlling their flow by electrostatic gating \cite{Yan2016, Dankert2017, Lin2017}. This should allow prompt exploration beyond our proof-of-principle system, MoTe$_2$, thereby accelerating the potential delivery of low-power spin-electronic devices and circuits \cite{Lin2019}. 
For example, the spin polarization generated by the \emph{canted} SHE can exert an out-of-plane antidamping torque in magnets with perpendicular magnetic anisotropy \cite{MacNeill2017, MacNeill2017PRB, Stiehl2019}, which are essential for next-generation, high-density spintronic applications \cite{Manchon2019, Liu2020a}. 

}

\section{Conclusions}

Our models and quantum transport calculations of the spintronic response of 1T$_\text{d}$ and 1T$^\prime$ MoTe$_2$ monolayers reveal the origin of a novel, canted SHE with long spin diffusion lengths, which reflects the unconventional spin textures allowed by their reduced symmetry and strong SOC. 
The obtained CSI figure of merit $\lambda_s \theta_{xy}\sim 1\text{--}50$ nm is superior to that of traditional spintronic materials (Pt, Au, W, and Ta) \cite{Isasa2015, Sinova2015} by up to two orders of magnitude. Given the similar electronic structures of MoTe$_2$ and WTe$_2$, including the persistent canted spin texture \cite{Garcia2020}, comparable performance is expected in the latter.
Our findings also call for a careful analysis of SHE measurements, since the interpretation of all-electrical detection in Hall bars \cite{Abanin2009, Song2020, Hankiewicz2004} usually ignores the possibility of multiple spin Hall components. We show how the presence of canted SHE can be experimentally identified by reciprocal SHE, and how the different SHC contributions may be isolated in a spin precession setup. 

\blue{Having precisely identified the underlying mechanism at play, these proof-of-principle results based on MoTe$_2$ suggest equally promising performance in several other identifiable material families with concurrent PST and large spin Berry curvature associated with low crystal symmetry and nontrivial electronic topology, respectively.

We finally mention that a much larger range of possible interesting materials should be available by engineering proximity effects and interfacial symmetries, as discussed for van der Waals heterostructures \cite{Sierra2021,Kurebayashi2021}.
}

\begin{acknowledgements}
M.V. acknowledges support from ``La Caixa'' Foundation and the Centre for Advanced 2D Materials at the National University of Singapore for its hospitality. X. W. acknowledges the ANR Flagera GRANSPORT funding. ICN2 authors were supported by the European Union Horizon 2020 research and innovation programme under Grant Agreement No. 881603 (Graphene Flagship) and  No. 824140 (TOCHA, H2020-FETPROACT-01-2018). ICN2 is funded by the CERCA Programme/Generalitat de Catalunya, and is supported by the Severo Ochoa program from Spanish MINECO (Grant No. SEV-2017-0706 and PID2019-111773RB-I00/ AEI / 10.13039/501100011033 ).  V.M.P. acknowledges the support of the National Research Foundation (Singapore) under its Medium-Sized Centre Programme (R-723-000-001-281).
\end{acknowledgements}

%********************references*************************************************
%BibTeX
%\bibliography{bibmote2} 

\begin{thebibliography}{94}
\bibitem [{foo()}]{footnote}%
  \BibitemOpen
  \href@noop {} {}\bibinfo {note} {In experiments, $R_\mathrm{ISHE}$ also
  depends on the TMD's electrical resistance \cite{Suppmat} and the Fermi
  level, as seen in Fig. \ref{fig_F3}. Interfacial barriers between graphene
  and the TMD can suppress the spin sink effect and cause a more distributed
  spin absorption across the TMD width, in which case the observed signal might
  be partially reduced.}\BibitemShut {Stop}%
\end{thebibliography}


\end{document}